\documentstyle{article}

\author{L.S.F.Olavo\\
Departamento de Fisica, Universidade de Brasilia,\\
70910-900, Brasilia-D.F., Brazil}
\title{Quantum Mechanics as a Classical Theory X:\\
Quantization in Generalized Coordinates
}

\begin{document}

\maketitle
\begin{abstract}
In this tenth paper of the series we aim at showing that our formalism,
using the Wigner-Moyal Infinitesimal Transformation together with classical
mechanics, endows us with the ways to quantize a system in any coordinate
representation we wish. This result is necessary if one even think about
making general relativistic extensions of the quantum formalism. Besides,
physics shall not be dependent on the specific representation we use and
this result is necessary to make quantum theory consistent and complete.
\end{abstract}

\section{Overview}

The present paper is the tenth part of a series of papers on the
mathematical and epistemological foundations of quantum theory. The papers
of this series may fit into one (or more) of four rather different
categories of interest.

They may be considered as reconstruction papers, where the existing
formalism is derived within the (classical) approach defined by the use of
Wigner-Moyal's Infinitesimal Transformation. Into such division we might put
papers I,II,III\cite{eu-1,eu-2,eu-3} and also papers VI,VII and VIII\cite
{eu-6,eu-7,eu-8}.

There are also papers where we resize the underlying epistemology to fit the
(purely classical) mathematical developments of the theory. In these cases,
alleged purely quantum effects are reinterpreted on classical grounds---from
the epistemological perspective. Into this category we might put papers
III,VI,VII and IX\cite{eu-3,eu-6,eu-7,eu-9}.

Pertaining to another class of papers we have those trying to expand the
applicability of quantum theory to other fields of investigation. It is
remarkable that such a task was taken with paper II\cite{eu-2}, where a
quantum theory (of one particle) that takes into account the effects of
gravity was developed, and with papers IV and V\cite{eu-4,eu-5}, where this
generalized relativistic quantum theory was applied to show that it predicts
particles with negative masses.

The last category is the one where we try to investigate the boundaries of
quantum theory---there where it seems to give unsatisfactory answers, both
from the mathematical and epistemological points of view. We might cite
paper IX as one example where the important question of operator formation
was discussed at length. Papers belonging to this category try to remove
from the theory some of its formal problems, as was the case with paper IX,
or misinterpretations, as was the case with paper III where the problem of
non-locality was investigated.

The present paper pertains to this last class of interest and is
particularly related with formal problems. We have been taught, since our
very introduction to the study of quantum mechanics, that, for quantizing a
system, we shall first write its {\it classical} hamiltonian in {\it %
cartesian coordinates}. This quantization may be mathematically represented
by 
\begin{equation}
\label{1}{\bf p}\rightarrow -i\hbar \nabla \ ,\ {\bf x}\rightarrow x, 
\end{equation}
where ${\bf p}$ is the momentum and ${\bf x}$ the coordinate, and by the
transformation 
\begin{equation}
\label{2}H({\bf p},{\bf x)}\rightarrow H(-i\hbar \nabla ,x). 
\end{equation}
It is only after this quantization has been performed that we may change to
another system of coordinates, distinct form the cartesian one\cite{Fock}.

This seems to be a terrible problem, although not seemed as such by many of
us, since physics is supposed to treat any (mathematical) system of
coordinates on the same grounds. If not for this reason, one may wonder
about the future of general relativistic extensions of a theory that {\it %
needs} the flat cartesian system of coordinates to exist. Within a
relativistic theory this need seems to be a scandal that denies {\it from
the very beginning} any such extension.

One may find in the literature \cite{Gruber1,Gruber2,Pauli} some trials to
overcome these difficulties, but even these approaches are permeated with
additional suppositions as in ref. \cite{Gruber1,Gruber2} where the author
has to postulate that the total quantum-mechanical momentum operator $%
p_{q_i} $ corresponding to the generalized coordinate $q_i$ is given by 
\begin{equation}
\label{2.1}p_{q_i}=-i\hbar \frac \partial {\partial q_i} 
\end{equation}
and has also to write the classical hamiltonian (the kinetic energy term) as 
\begin{equation}
\label{2.2}H=\frac 1{2m}\sum_{i,k}p_{q_i}^{*}g^{ik}p_{q_k} 
\end{equation}
whatever be the complex conjugate of the classical momentum {\it function}.

These approaches seem to be rather unsatisfactory for we would like to
derive our results using only first principles, without having to add more
postulates to the theory.

This problem appears because quantum mechanics, as developed in textbooks,
is not a theory with a clearly discernible set of axioms\cite{Mehra}.
Indeed, the rules (\ref{1}) and (\ref{2}) above are part of the fuzzy set of
axioms one could append to it.

We have developed a completed axiomatic (classical) version of quantum
mechanics which, we expect, does not depend on the specific set of
coordinates used.

The aim of this paper is to show that our expectations are confirmed by the
mathematical formalism.

We will show in the second section how to quantize a hamiltonian with a
central potential in spherical coordinates using only the three axioms we
have already postulated, now written in this coordinate system. This will
serve as an illustrative example of how quantization in generalized
coordinates shall be done.

The third section will aim at generalizing the previous particular approach
to any set of orthogonal generalized coordinates; that is, to show how to
quantize in such coordinates.

\section{Spherical Coordinates: an example}

We begin by rewriting our three axioms\cite{eu-1} in the appropriate
coordinate system:

\begin{description}
\item[Axiom 1:]  Newtonian particle mechanics is valid for all particles
which constitute the systems composing the {\it ensemble};

\item[Axiom 2:]  For an {\it ensemble} of isolated systems the joint
probability density function is a constant of motion: 
\begin{equation}
\label{3}\frac{dF(r,\theta ,\phi ,p_r,p_\theta ,p_\phi ;t)}{dt}=0;
\end{equation}

\item[Axiom 3:]  The Wigner-Moyal Infinitesimal Transformation, defined as 
\begin{equation}
\label{4}\rho ({\bf r}-\frac{\delta {\bf r}}2,{\bf r+}\frac{\delta {\bf r}}%
2;t)=\int F(r,\theta ,\phi ,p_r,p_\theta ,p_\phi ;t)\exp (\frac i\hbar {\bf p%
}\cdot \delta {\bf r})d^3p
\end{equation}
may be applied to represent the dynamics of the system in terms of functions 
$\rho ({\bf r},\delta {\bf r};t)$.
\end{description}

Using expression (\ref{3}) (Liouville's equation), we may write 
\begin{equation}
\label{5}\frac{\partial F}{\partial t}+\{H,F\}=0, 
\end{equation}
where $H$ is the hamiltonian and $\{,\}$ is the classical Poisson bracket.
By means of the hamiltonian, written in spherical coordinates 
\begin{equation}
\label{6}H=\frac 1{2m}\left[ p_r^2+\frac{p_\theta ^2}{r^2}+\frac{p_\phi ^2}{%
r^2\sin {}^2(\theta )}\right] +V(r), 
\end{equation}
we find the Liouville equation%
$$
\frac{\partial F}{\partial t}+\frac{p_r}m\frac{\partial F}{\partial r}+\frac{%
p_\theta }{mr^2}\frac{\partial F}{\partial \theta }+\frac{p_\phi }{mr^2\sin
{}^2(\theta )}\frac{\partial F}{\partial \phi }- 
$$
\begin{equation}
\label{7}-\left[ \frac{\partial V}{\partial r}-\frac{p_\theta ^2}{mr^3}- 
\frac{p_\phi ^2}{mr^3\sin {}^2(\theta )}\right] \frac{\partial F}{\partial
p_r}+\frac{p_\phi ^2}{mr^2\sin {}^2(\theta )}\cot (\theta )\frac{\partial F}{%
\partial p_\theta }=0. 
\end{equation}

As a means of writing the Wigner-Moyal Transformation in spherical
coordinates, we note that 
\begin{equation}
\label{8}\delta {\bf r}=\delta r\widehat{{\bf r}}+r\delta \theta \widehat{%
{\bf \theta }}+r\sin (\theta )\delta \phi \widehat{{\bf \phi }} 
\end{equation}
where $(\widehat{{\bf r}},\widehat{{\bf \theta }},\widehat{{\bf \phi }})$
are the unit normals, and 
\begin{equation}
\label{9}{\bf p}=p_r\widehat{{\bf r}}+\frac{p_\theta }r\widehat{{\bf \theta }%
}+\frac{p_\phi }{r\sin (\theta )}\widehat{{\bf \phi }}, 
\end{equation}
giving 
\begin{equation}
\label{10}{\bf p}\cdot \delta {\bf r}=\delta r\cdot p_r+\delta \theta \cdot
p_\theta +\delta \phi \cdot p_\phi . 
\end{equation}

Using the relation between the unit normals in cartesian $({\bf i},{\bf j},%
{\bf k})$ and spherical $(\widehat{{\bf r}},\widehat{{\bf \theta }},\widehat{%
{\bf \phi }})$ coordinates 
\begin{equation}
\label{11}\left\{ 
\begin{array}{l}
\widehat{{\bf r}}={\bf i}\sin (\theta )\cos (\phi )+{\bf j}\sin (\theta
)\sin (\phi )+{\bf k}\cos (\theta )\  \\ \widehat{{\bf \theta }}={\bf i}\cos
(\theta )\cos (\phi )+{\bf j}\cos (\theta )\sin (\phi )-{\bf k}\sin (\theta
) \\ \widehat{{\bf \phi }}=-{\bf i}\sin (\phi )+{\bf j}\cos (\phi ) 
\end{array}
\right. , 
\end{equation}
we find the following relation between the momenta: 
\begin{equation}
\label{12}\left\{ 
\begin{array}{l}
p_x=p_r\sin (\theta )\cos (\phi )+(p_\theta /r)\cos (\theta )\cos (\phi
)-(p_\phi /r)(\sin (\phi )/\sin (\theta ))\  \\ 
p_y=p_r\sin (\theta )\sin (\phi )+(p_\theta /r)\cos (\theta )\sin (\phi
)+(p_\phi /r)(\cos (\phi )/\sin (\theta ))\  \\ 
p_z=p_r\cos (\theta )-(p_\theta /r)\sin (\theta ) 
\end{array}
\right. . 
\end{equation}
The Jacobian relating the two infinitesimal volume elements 
\begin{equation}
\label{13}dp_xdp_ydp_z=\left\| J\right\| _pdp_rdp_\theta dp_\phi 
\end{equation}
is given by 
\begin{equation}
\label{14}\left\| J\right\| _p=\frac 1{r^2\sin (\theta )}. 
\end{equation}

It is now possible to rewrite expression (\ref{4}) as 
\begin{equation}
\label{15}\rho ({\bf r}-\frac{\delta {\bf r}}2,{\bf r+}\frac{\delta {\bf r}}%
2;t)=\int F({\bf r},{\bf p};t)e^{\frac i\hbar (\delta r\cdot p_r+\delta
\theta \cdot p_\theta +\delta \phi \cdot p_\phi )}\frac{dp_rdp_\theta
dp_\phi }{r^2\sin (\theta )}.
\end{equation}

With equation (\ref{7}) and expression (\ref{15}) at hands we may find the
equation satisfied by the density function $\rho ({\bf r},\delta {\bf r};t)$
in exactly the same way as was previously done for cartesian coordinates\cite
{eu-1}---it is noteworthy that now we have the jacobian (\ref{14}) that will
change slightly the appearance of this equation terms.

After some straightforward calculations we arrive at%
$$
-\frac{\hbar ^2}m\left[ \frac 1{r^2}\frac \partial {\partial r}\left( r^2
\frac{\partial \rho }{\partial (\delta r)}\right) +\frac 1{r^2\sin (\theta
)}\frac \partial {\partial \theta }\left( \sin (\theta )\frac{\partial \rho 
}{\partial (\delta \theta )}\right) +\frac 1{r^2\sin {}^2(\theta )}\frac{%
\partial ^2\rho }{\partial \phi \partial (\delta \phi )}\right] + 
$$
\begin{equation}
\label{16}+\frac{\hbar ^2}m\left[ \frac{\delta r}{r^3}\frac{\partial ^2\rho 
}{\partial (\delta \theta )^2}+\frac{\delta r}{r^3\sin {}^2(\theta )}\frac{%
\partial ^2\rho }{\partial (\delta \phi )^2}+\frac{\delta \theta \cot
(\theta )}{r^2\sin {}^2(\theta )}\frac{\partial ^2\rho }{\partial (\delta
\phi )^2}\right] +\delta r\frac{\partial V}{\partial r}\rho =i\hbar \frac{%
\partial \rho }{\partial t}.
\end{equation}

To go from this equation to the equation for the amplitudes we may write 
\begin{equation}
\label{17}\rho ({\bf x}-\frac{\delta {\bf x}}2,{\bf x+}\frac{\delta {\bf x}}%
2;t)=\psi ^{\dagger }({\bf x}-\frac{\delta {\bf x}}2;t)\psi ({\bf x+}\frac{%
\delta {\bf x}}2;t)
\end{equation}
and also write 
\begin{equation}
\label{18}\psi ({\bf x};t)=R({\bf x};t)\exp \left( iS({\bf x};t)/\hbar
\right) 
\end{equation}
in cartesian coordinates, for example. We then expand expression (\ref{17})
around the infinitesimals quantities to get, until second order,%
$$
\rho ({\bf x},\delta {\bf x};t)=\left\{ R^2+\frac R4\sum_{i,j=1}^3\delta
x_i\delta x_j\frac{\partial ^2R}{\partial x_i\partial x_j}-\frac
14\sum_{i,j=1}^3\delta x_i\delta x_j\frac{\partial R}{\partial x_i}\frac{%
\partial R}{\partial x_j}\right\} \cdot  
$$
\begin{equation}
\label{19}\cdot \exp \left[ \frac i\hbar \left( \delta x\frac{\partial S}{%
\partial x}+\delta y\frac{\partial S}{\partial y}+\delta z\frac{\partial S}{%
\partial z}\right) \right] ,
\end{equation}
where $x_i,i=1,2,3$ implies $x,y,z$, and use the relations 
\begin{equation}
\label{20}\left\{ 
\begin{array}{l}
\partial _x=\sin (\theta )\cos (\phi )\partial _r+(1/r)\cos (\theta )\cos
(\phi )\partial _\theta -(1/r)(\sin (\phi )/\sin (\theta ))\partial _\phi \ 
\\ 
\partial _y=\sin (\theta )\sin (\phi )\partial _r+(1/r)\cos (\theta )\sin
(\phi )\partial _\theta +(1/r)(\cos (\phi )/\sin (\theta ))\partial _\phi \ 
\\ 
\partial _z=\cos (\theta )\partial _r-(1/r)\sin (\theta )\partial _\theta 
\end{array}
\right. ,
\end{equation}
where $\partial _u$ is an abbreviation of $\partial /\partial u$, to write 
the density function in spherical coordinates as%
$$
\rho ({\bf r},\delta {\bf r};t)=\left\{ R^2+\frac R4\left[ \delta r^2\frac{%
\partial ^2R}{\partial r^2}+\delta \theta ^2\left( \frac{\partial ^2R}{%
\partial \theta ^2}+r\frac{\partial R}{\partial r}\right) +\right. \right.  
$$
$$
\left. +\delta \phi ^2\left( \frac{\partial ^2R}{\partial \phi ^2}+r\sin
{}^2(\theta )\frac{\partial R}{\partial r}+\cos (\theta )\sin (\theta )\frac{%
\partial R}{\partial \theta }\right) +\right.  
$$
$$
\left. 2\delta r\delta \theta \left( \frac{\partial ^2R}{\partial r\partial
\theta }-\frac 1r\frac{\partial R}{\partial \theta }\right) +2\delta r\delta
\phi \left( \frac{\partial ^2R}{\partial r\partial \phi }-\frac 1r\frac{%
\partial R}{\partial \phi }\right) -\right.  
$$
$$
\left. \left. +2\delta \theta \delta \phi \left( \frac{\partial ^2R}{%
\partial \theta \partial \phi }-\cot (\theta )\frac{\partial R}{\partial
\phi }\right) \right] -\frac 14\sum_{i,j=1}^3\delta x_i\delta x_j\frac{%
\partial R}{\partial x_i}\frac{\partial R}{\partial x_j}\right\} \cdot  
$$
\begin{equation}
\label{21}\cdot \exp \left[ \frac i\hbar \left( \delta r\frac{\partial S}{%
\partial r}+\delta \theta \frac{\partial S}{\partial \theta }+\delta \phi 
\frac{\partial S}{\partial \phi }\right) \right] ,
\end{equation}
where now $x_i,i=1,2,3$ means $r,\theta ,\phi $.

The next step is to take expression (\ref{21}) into equation (\ref{16}) and
to collect the zeroth and first order terms in the infinitesimals%
\footnote{this cumbersome exercise was performed using algebraic computation.}
to get, as usual\cite{eu-1}, the equations (written in spherical
coordinates) 
\begin{equation}
\label{22}\delta {\bf r}\cdot \frac \partial {\partial {\bf r}}\left[ \frac{%
\partial S}{\partial t}+\frac 1{2m}\left( \nabla S\right) ^2+V(r)-\frac{%
\hbar ^2}{2mR}\nabla ^2R\right] =0 
\end{equation}
and 
\begin{equation}
\label{23}\frac{\partial R^2}{\partial t}+\nabla \cdot \left( \frac{R^2}%
m\nabla S\right) =0. 
\end{equation}

Equation (\ref{23}) is the continuity equation while equation (\ref{22}) is
the same as writing 
\begin{equation}
\label{24}\delta {\bf r}\cdot \frac \partial {\partial {\bf r}}\left[ \left( 
\widehat{H}\psi -i\hbar \frac{\partial \psi }{\partial t}\right) \frac 1\psi
\right] =0, 
\end{equation}
where the $\psi $ is as given in expression (\ref{18}) and the hamiltonian
operator $H$ is given by 
\begin{equation}
\label{25}\widehat{H}=\frac{-\hbar ^2}{2m}\left[ \frac 1{r^2}\frac \partial
{\partial r}\left( r^2\frac \partial {\partial r}\right) +\frac 1{r^2\sin
(\theta )}\frac \partial {\partial \theta }\left( \sin (\theta )\frac
\partial {\partial \theta }\right) +\frac 1{r^2\sin {}^2(\theta )}\frac{%
\partial ^2}{\partial \phi ^2}\right] +V(r). 
\end{equation}

In this sense, we may say, since the infinitesimals are all independent,
that we have derived the Schr\"odinger equation written as%
$$
\frac{-\hbar ^2}{2m}\left[ \frac 1{r^2}\frac \partial {\partial r}\left(
r^2\frac \partial {\partial r}\right) +\frac 1{r^2\sin (\theta )}\frac
\partial {\partial \theta }\left( \sin (\theta )\frac \partial {\partial
\theta }\right) +\frac 1{r^2\sin {}^2(\theta )}\frac{\partial ^2}{\partial
\phi ^2}\right] \psi + 
$$
\begin{equation}
\label{26}+V(r)\psi =i\hbar \frac{\partial \psi }{\partial t},
\end{equation}
where the constant coming from expression (\ref{24}) might be appended in
the right hand term above and reflects a mere definition of a new reference
energy level.

We thus have quantized the system using only spherical coordinates from the
very beginning as was our interest to show.

In the next section we will generalize this result to any set of orthogonal
coordinate systems.

\section{Orthogonal Coordinates}

In this case we have the transformation rules 
\begin{equation}
\label{27}x_\alpha =x_\alpha (u_1,u_2,u_3)\ ;\ \alpha =1,2,3, 
\end{equation}
where $x_\alpha $ are the coordinates written in some system\ (not
necessarily the cartesian one), $u_i$ are the coordinates in the new system
and the differential line element is given by\cite{Gradystein} 
\begin{equation}
\label{28}d{\bf r}=h_1du_1{\bf e}_1+h_2du_2{\bf e}_2+h_3du_3{\bf e}_3, 
\end{equation}
where the ${\bf e}_i$ are the unit normals in the new $(u_1,u_2,u_3)$%
-coordinate system and 
\begin{equation}
\label{29}h_i{\bf e}_i=\frac{\partial {\bf r}}{\partial u^i}. 
\end{equation}

The momenta $(p_1,p_2,p_3)$ canonically conjugate to the $u$-coordinate
system are given by%
$$
{\bf p}=m\left( h_1\frac{du_1}{dt}{\bf e}_1+h_2\frac{du_2}{dt}{\bf e}_2+h_3
\frac{du_3}{dt}{\bf e}_3\right) = 
$$
\begin{equation}
\label{30}=\left( \frac{p_1}{h_1}{\bf e}_1+\frac{p_2}{h_2}{\bf e}_2+\frac{p_3
}{h_3}{\bf e}_3\right) ,
\end{equation}
or 
\begin{equation}
\label{30.a}p_i=mh_i^2\frac{du_i}{dt},
\end{equation}
such that 
\begin{equation}
\label{31}{\bf p}\cdot \delta {\bf u}=p_1\delta u_1+p_2\delta u_2+p_3\delta
u_3.
\end{equation}

The Hamiltonian may be written as\cite{Brillouin} 
\begin{equation}
\label{32}H=\frac 1{2m}\left[ \frac{p_1^2}{h_1^2}+\frac{p_2^2}{h_2^2}+\frac{%
p_3^2}{h_3^2}\right] +V({\bf u}) 
\end{equation}
and the Liouville equation becomes 
\begin{equation}
\label{33}\frac{\partial F}{\partial t}+\sum_{i=1}^3\frac{p_i^2}{mh_i^2} 
\frac{\partial F}{\partial u_i}+\sum_{j=1}^3\left[ \sum_{i=1}^3\left( \frac{%
p_i^2}{mh_i^3}\frac{\partial h_i}{\partial u_j}\right) -\frac{\partial V}{%
\partial u_j}\right] \frac{\partial F}{\partial p_j}=0. 
\end{equation}

Since the coordinate transformation (\ref{27}) is a special type of
canonical transformation, we shall have the Jacobian of the momentum
transformation given by 
\begin{equation}
\label{34}\left\| J\right\| _p=\frac 1{h_1h_2h_3}, 
\end{equation}
for the Jacobian of the coordinate transformation is 
\begin{equation}
\label{35}\left\| J\right\| _u=h_1h_2h_3 
\end{equation}
and the infinitesimal phase space volume element is a canonical invariant%
\cite{Goldstein}.

With the density function given by 
\begin{equation}
\label{36}\rho \left( {\bf u}-\frac{\delta {\bf u}}2,{\bf u+}\frac{\delta 
{\bf u}}2;t\right) =\int F({\bf u},{\bf p};t)\exp \left( \frac i\hbar {\bf %
p\cdot }\delta {\bf u}\right) \frac{dp_1dp_2dp_3}{h_1h_2h_3},
\end{equation}
it is straightforward to find the differential equation it satisfies as%
$$
-i\hbar \frac{\partial \rho }{\partial t}-\frac{\hbar ^2}m\left\{ \frac
1{h_1h_2h_3}\left[ \frac \partial {\partial u_1}\left( \frac{h_2h_3}{h_1}%
\frac \partial {\partial (\delta u_1)}\right) +\frac \partial {\partial
u_2}\left( \frac{h_1h_3}{h_2}\frac \partial {\partial (\delta u_2)}\right)
+\right. \right.  
$$
\begin{equation}
\label{37}\left. \left. +\frac \partial {\partial u_3}\left( \frac{h_1h_2}{%
h_3}\frac \partial {\partial (\delta u_3)}\right) \right] -\sum_{i,j=1}^3
\frac{\delta u_i}{h_j^3}\frac{\partial h_j}{\partial u_i}\frac{\partial
^2\rho }{\partial (\delta u_i)^2}\right\} +\sum_{i=1}^3\delta u_i\frac{%
\partial V}{\partial u_i}\rho =0.
\end{equation}
The reader may easily verify that, with the coordinate transformation given
by 
\begin{equation}
\label{38}\left\{ 
\begin{array}{l}
x=r\sin \theta \cos \phi  \\ 
y=r\sin \theta \sin \phi  \\ 
z=r\cos \theta 
\end{array}
\right. \ ;\ \left\{ 
\begin{array}{l}
h_1=1 \\ 
h_2=r \\ 
h_3=r\sin \theta 
\end{array}
\right. ,
\end{equation}
we recover the result of equation (\ref{16}).

We now write the density function as 
\begin{equation}
\label{39}\rho ({\bf u}-\frac{\delta {\bf u}}2,{\bf u+}\frac{\delta {\bf u}}%
2;t)=\psi ^{\dagger }({\bf u}-\frac{\delta {\bf u}}2;t)\psi ({\bf u+}\frac{%
\delta {\bf u}}2;t), 
\end{equation}
with the, generally complex, amplitudes written as 
\begin{equation}
\label{40}\psi ({\bf x};t)=R({\bf u};t)\exp \left( iS({\bf u};t)/\hbar
\right) 
\end{equation}
and expand these amplitudes until second order in the infinitesimal
parameter $\delta {\bf u}$, to find (until second order)%
$$
\rho ({\bf u},\delta {\bf u};t)=\left\{ R^2+\frac R4\left[
\sum_{i,j=1}^3\delta u_i\delta u_j\left( \frac{\partial ^2R}{\partial
u_i\partial u_j}-\sum_{k=1}^3\Gamma _{ij}^k\frac{\partial R}{\partial u_k}%
\right) \right] -\right. 
$$
\begin{equation}
\label{41}\left. -\frac 14\left[ \sum_{i,j=1}^3\delta u_i\delta u_j\frac{%
\partial R}{\partial u_i}\frac{\partial R}{\partial u_j}\right] \right\}
\cdot \exp \left( \frac i\hbar \sum_{i=1}^3\delta u_i\frac{\partial S}{%
\partial u_i}\right) , 
\end{equation}
where $\Gamma $ is the Christoffel Symbol\cite{Weinberg}. The reader may
verify that expression (\ref{41}) gives the correct (\ref{21}) result when
expressed in spherical coordinates.

We then insert this expression into equation (\ref{37}) and collect the
zeroth and first order terms on the infinitesimals%
\footnote{Again, algebraic computation was used throughout.} to get the
equations (written in general orthogonal coordinates) 
\begin{equation}
\label{42}\delta {\bf u}\cdot \frac \partial {\partial {\bf u}}\left[ \frac{%
\partial S}{\partial t}+\frac 1{2m}\left( \nabla S\right) ^2+V({\bf u})- 
\frac{\hbar ^2}{2mR}\nabla ^2R\right] =0 
\end{equation}
and 
\begin{equation}
\label{43}\frac{\partial R^2}{\partial t}+\nabla \cdot \left( \frac{R^2}%
m\nabla S\right) =0. 
\end{equation}

Equation (\ref{43}) is the continuity equation while equation (\ref{42}) is
the same as writing 
\begin{equation}
\label{44}\delta {\bf u}\cdot \frac \partial {\partial {\bf u}}\left[ \left( 
\widehat{H}\psi -i\hbar \frac{\partial \psi }{\partial t}\right) \frac 1\psi
\right] =0,
\end{equation}
where the $\psi $ is as shown in expression (\ref{40}) and the hamiltonian
operator $\widehat{H}$ is given by 
$$
\widehat{H}=\frac{-\hbar ^2}{2m}\frac 1{h_1h_2h_3}\left[ \frac \partial
{\partial u_1}\left( \frac{h_2h_3}{h_1}\frac \partial {\partial u_1}\right)
+\frac \partial {\partial u_2}\left( \frac{h_2h_3}{h_1}\frac \partial
{\partial u_2}\right) \right.  
$$
\begin{equation}
\label{45}\left. +\frac \partial {\partial u_3}\left( \frac{h_1h_2}{h_3}%
\frac \partial {\partial u_3}\right) \right] +V({\bf u}).
\end{equation}

Because the infinitesimals are all independent, the term inside brackets in
equation (\ref{44}) must vanish identically. This allows us to say that we
have derived the Schr\"odinger equation written as%
$$
\frac{-\hbar ^2}{2m}\frac 1{h_1h_2h_3}\left[ \frac \partial {\partial
u_1}\left( \frac{h_2h_3}{h_1}\frac \partial {\partial u_1}\right) +\frac
\partial {\partial u_2}\left( \frac{h_2h_3}{h_1}\frac \partial {\partial
u_2}\right) \right.  
$$
\begin{equation}
\label{46}\left. +\frac \partial {\partial u_3}\left( \frac{h_1h_2}{h_3}%
\frac \partial {\partial u_3}\right) \right] \psi +V({\bf u)}\psi =i\hbar 
\frac{\partial \psi }{\partial t},
\end{equation}
where the constant coming from expression (\ref{44}) might be appended in
the right hand term above and reflects a mere new definition of the
reference energy level---or a new definition of the function $S$ by adding
this constant factor.

We thus have quantized the system using only general orthogonal coordinates
from the very beginning as was our aim to show. The extension of this result
for non-orthogonal coordinate systems is straightforward and will not be
done here.

\section{Conclusions}

We have shown that the process of quantization is coordinate system
independent. This result does not bring anything new to the machinery of
quantum theory, since its formal apparatus for application on computational
problems remains untouched.

However, when showing this coordinate system independence, we also show that
the formalism is in conformity with our expectations that {\it physics shall
not depend on the way we choose to represent it}.

Although this seems to be sterile from the point of view of calculations, it
gives the formalism coherence and wideness. Coherence for the reasons
explained above and wideness for now it is possible to justify any trial to
find a general relativistic extension of this formalism.

These results may also be seen as another confirmation of our guesses about
the classical nature of quantum theory.

\end{document}